\documentclass[aps,prb,twocolumn]{revtex4}
\usepackage{graphicx}
\usepackage{amsmath}

\def\vec#1{{\bf #1}}

\newcommand{\pLJ}{\psi_{LJ}}

\begin{document}

\title{A $\nu=2/5$ Paired Wavefunction}

\author{Steven H. Simon${}^a$, E. H. Rezayi${}^{b}$,  N. R. Cooper${}^c$, and I. Berdnikov${}^{d,a}$}

\affiliation{${}^a$ Lucent Technologies, Bell Labs, Murray Hill,
NJ, 07974
\\${}^b$Department of Physics, California State University,
 Los Angeles California 90032
\\${}^c$  Department of Physics, Cavendish Laboratory, J J Thomson Avenue, Cambridge
CB3 0HE
\\${}^d$ Department of Physics, Rutgers University,  136
Frelinghuysen Rd., Piscataway NJ 08854}

\date{August 15, 2006}

\begin{abstract}
We construct a wavefunction, generalizing the well known
Moore-Read Pfaffian, that describes spinless electrons at filling
fraction $\nu=2/5$ (or bosons at filling fraction $\nu=2/3$) as
the ground state of a very simple three body potential.  We find,
analogous to the Pfaffian, that when quasiholes are added there is
a ground state degeneracy which can be identified as zero-modes of
the quasiholes.   The zero-modes are identified as having semionic
statistics.  We write this wavefunction as a correlator of the
 Virasoro minimal model conformal field theory ${\cal
M}(5,3)$.   Since this model is non-unitary, we conclude that this
wavefunction is a quantum critical state.   Nonetheless, we find
that the overlaps of this wavefunction with exact diagonalizations
in the lowest and first excited Landau level are very high,
suggesting that this wavefunction may have experimental relevance
for some transition that may occur in that regime.
\end{abstract}

\maketitle

\section{Introduction}

The vast majority of quantum Hall states observed experimentally in
the lowest Landau level (LLL) are very accurately described in terms
of composite fermion\cite{Olle} (or equivalently the
hierarchy\cite{Prange}) wavefunctions. Despite these successes,
there are a number of quantum Hall states that remain much less well
understood and may require more exotic explanations. For example,
there exist a few experimentally observed quantum Hall plateaus in
the LLL that do not fit into the usual hierarchy-composite-fermion
framework\cite{Pan}. Also, in the first excited Landau level (1LL),
even the ``simple" observed filling fractions (such as $\nu=2+1/3$
and $2+2/5$) appear from numerics to have strong differences from
the corresponding states in the LLL\cite{Prange,Quinn,Edunpub}.  Of
the non-hierarchy exotic states that have been proposed, perhaps the
best understood is the Moore-Read Pfaffian\cite{Moore}, which is
thought to describe\cite{Pfaff} the plateau observed at $\nu=2+1/2$,
and whose quasiparticle excitations have exotic nonabelian
statistics. However, for the neighboring experimentally observed
plateau\cite{Xia} at $\nu=2+2/5$ there are at least two competing
trial states which have been proposed: the hierarchy (composite
fermion) state \cite{Olle,Prange}, and the (particle-hole conjugate
of the) $Z_3$ Read-Rezayi parafermion
state\cite{ReadRezayi,EdandNick}--- a generalization of the Pfaffian
which has an even richer nonabelian structure.  Another case where
more exotic states may occur is in the quantum Hall effect of
rotating bosons\cite{Bosons}.

In this paper we will study another type of generalization of the
Pfaffian that gives a different trial state at $\nu=2/5$ (or $2 +
2/5$) for spinless electrons.  We call this new wavefunction the
``Gaffnian", for reasons described below.  The Hamiltonian that
generates this state as its unique highest density zero energy
state is an extremely simple generalization of the Hamiltonian
that similarly generates the Moore-Read Pfaffian (and is also
similar in spirit to the Hamiltonians that generate the
Read-Rezayi wavefunctions). Similar to the Pfaffian and
Read-Rezayi states, when additional flux is added, there is a
degeneracy of states associated with zero-modes of the quasiholes.
In the present case the zero-modes have semionic statistics,
compared to fermionic statistics in the Pfaffian case, or
parafermionic statistics in the Read-Rezayi case.  We then write
this wavefunction as the correlator of a conformal field theory,
known as the Virasoro ${\cal M}(5,3)$ minimal model.   Since this
field theory is nonunitary we conclude that the wavefunction does
not represent a phase, but rather a quantum critical point between
phases.  As such, the concept of nonabelian statistics is not
necessarily applicable.  Nonetheless, exact diagonalizations show
extremely high overlaps with this trial wavefunction.   We take
this to suggest that this wavefunction is likely relevant to a
phase transition that is somehow ``near" the hierarchy phase.

The outline of this paper is as follows.  In section
\ref{sec:Gaffnian} we will introduce this new wavefunction and
explore some of its properties.  We define this state as the
unique highest density zero energy state of a simple Hamiltonian.
We then consider what happens when additional flux is added to the
system.  As mentioned above, in the presence of quasiholes, there
is a ground state degeneracy stemming from semionic zero-modes.
Since some of the analytic manipulations are messy, we relegate
some of the details to rather lengthy appendices.  In particular,
the demonstration that the Gaffnian is a unique ground state of
this Hamiltonian, and the explicit counting of quasihole states is
put in appendix \ref{app:counting}. However, for the interested
reader, this appendix shows explicitly the mechanism by which the
semions occur.   In section \ref{sec:conformal} we construct the
Gaffnian as a correlator of the ${\cal M}(5,3)$ Virasoro minimal
model conformal field theory. In section \ref{sec:exact} we
examine results of exact diagonalizations.  We look at low energy
excitations to find evidence of criticality, and we also discover
that the overlap of the Gaffnian with the hierarchy wavefunction
is remarkably high (we also find high overlap with exact
diagonalizations of systems with interactions close to Coulomb).
Finally, in section \ref{sec:discussion} we give a brief
discussion of some of our results.

\section{The Gaffnian Wavefunction}
\label{sec:Gaffnian}

Before constructing our new trial wavefunction, for motivation we
review construction of Laughlin wavefunctions. Constrained to a
single Landau level (LLL or 1LL), the relative angular momentum of
two fermions $L_2$ must be odd and positive. Thus the minimum
relative angular momentum is $L_{2}^{min}=1$ (For bosons,
$L_2^{min} = 0$ and $L_2$ must be even).   We define a projection
operator $P_{2}^p$ to project out (i.e, to keep only) states where
any two particles have relative angular momentum less than
$L^{min}_2+p$ (with $p\,$ even). This projection
operator\cite{Haldane} can serve as a Hamiltonian. The Laughlin
$\nu=1/(L^{min}_2 + p)$ state is the unique highest density (zero
energy) ground state of $P_2^p$ (with $L_2^{min}+p$ odd for
fermions and even for bosons).

We now generalize this construction.  In a single Landau level the
relative angular momentum of {\it three} fermions $L_3$ has a
minimum value $L_3^{min}=3$ (For bosons, $L_3^{min}=0$).   It is
not hard to show\cite{ReadRezayiPfaff,Green,Us} that $L_3 \neq
L_3^{min}+1$ is dictated by symmetry, but any other $L_3 \geq
L_3^{min}$ is allowed.   We analogously define a projection
operator $P_{3}^p$ to project out (i.e., to keep only) states
where any three particles have relative angular momentum less than
$L_3^{min} + p$ which will serve as our Hamiltonian.   It is well
known\cite{Moore,ReadRezayiPfaff,Greiter} that the Pfaffian (at
$\nu=1/2$ for fermions and $\nu=1$ for bosons) is the unique
highest density (zero energy) ground state of the Hamiltonian
$P_3^2$.  In Ref.~\onlinecite{Green} another state, known as the
``Haffnian" is shown to be the unique highest density (zero
energy) ground state of $P_3^4$ (which is a $\nu=1/3$ state for
fermions and $\nu=1/2$ state for bosons). Using the method of
\onlinecite{ReadRezayiPfaff,Green} we can show that the
Hamiltonian $P_3^3$ also has a unique highest density (zero
energy) ground state.  The argument is straightforward and is
given in the appendix of this paper. This unique state occurs at
$\nu=2/5$ for fermions ($\nu=2/3$ for bosons), and is the focus of
this paper. Since this new $p=3$ state lies between the $p=2$
P{\bf F}affian and the $p=4$ {\bf H}affnian, we alpha-phonetically
interpolate and dub this new state the ``{\bf G}affnian".

Before commencing our study of the Gaffnian, we note that several
other states can be constructed analogously.  By considering
general $k$-particle angular momenta $L_k$, we can construct a
general $P_k^p$. In Ref.~\onlinecite{ReadRezayi} it was shown that
the Hamiltonian $P_k^2$ generates the $Z_{k-1}$ Read-Rezayi state
(the $Z_2$ state being the Pfaffian). Study of several other
values of $p$ and $k$ is given in Ref. \onlinecite{Us} by three of
the current authors.

Knowing that a unique quantum Hall ground state exists for the
Hamiltonian $P^3_3$, we set about describing the properties of the
Gaffnian.  We begin by writing down the wavefunction explicitly.

We will represent a particle's coordinate as an analytic variable
$z= x + iy$ which is simply the complex representation of the
particle position $\vec r$.   On the spherical geometry, $z$ is
the stereographic projection of the position on the sphere of
radius $R$ to the plane.  All distances are measured in units of
the magnetic length.  We can write any single particle
wavefunction as an analytic function times a measure $\mu(\vec
r)$. On the disk the measure is the usual gaussian
factor\cite{Prange} $  \mu(\vec r) = e^{-|z|^2/4}$ whereas on the
sphere\cite{Haldane} (with stereographic projection to the plane)
the measure is $
    \mu(\vec r) = [1 +  |z|^2 / (4 R^2)]^{-(1 + N_{\phi}/2)}
$  with $N_{\phi}$ (=$2 R^2$ when the magnetic length is unity)
being the total number of flux penetrating the
sphere\cite{ReadRezayi,Us}. On the sphere the degree of the
polynomial $\psi(z)$ ranges from $z^0$ to $z^{N_{\phi}}$ giving a
complete basis of the $N_{\phi} +1$ states of the LLL.  We will
not write the measure explicitly, instead writing all
wavefunctions simply as analytic functions (which must be fully
symmetric for bosons and fully antisymmetric for fermions).

It is convenient to think, for a moment about bosons at $\nu=2/3$.
Since the Hamiltonian $P^3_3$ puts no restriction on the two
particle angular momentum, there is no restriction against two
bosons being at the same point $z_0$. However, when a third
particle approaches, it must approach the other two\cite{Us} such
that the overall angular momentum of the three particles is $p
\geq 3$, i.e.,  the wavefunction vanishes as $(z_3 - z_0)^p$. (In
this sense, the Gaffnian, like the Pfaffian and Haffnian, is a
paired state in the spirit of that originally proposed in Ref.
\onlinecite{Halperin}.)  The Gaffnian wavefunction can be written
explicitly as\cite{endnote2}
\begin{widetext}
\begin{eqnarray}
    \label{eq:gaffwf}
        \Psi &=& \tilde S\left[\prod_{a < b \leq N/2} \!\!\!\!(z_a -
        z_b)^{2+q}
        \prod_{N/2 < c < d }  \!\!\!\! (z_c - z_d)^{2+q}  \prod_{e \leq N/2 < f }  \!\!\!\! (z_e - z_f)^{1+q}
\prod_{g \leq N/2} \frac{1}{(z_g - z_{g+N/2})} \right]
    \end{eqnarray}
    \end{widetext}
where $q=0$ corresponds to a bosonic ($\nu=2/3$) wavefunction and
$q=1$ is a fermionic ($\nu=2/5$) wavefunction.   We have assumed
the number $N$ of particles is even and $\tilde S$ means
symmetrize or antisymmetrize over all particle coordinates for
bosons or fermions respectively.  One can confirm directly that
the above wavefunction for $q=0$ correctly has the property that
it does not vanish as two particles come to the same position but
vanishes as three powers as the third particle arrives.  Further,
we show in appendix \ref{app:counting} that this is the unique
densest wavefunction that has this property.   As is standard in
spherical geometry\cite{Prange,Haldane} we can obtain the value of
the flux by looking at the maximum power of $z_i$ that occurs.
Counting powers of $z$ we find that the Gaffnian wavefunction
occurs on the sphere for total flux
\begin{equation}
\label{eq:flux} N_{\phi} = 3 N/2 -3 + q(N-1).
\end{equation}
 This value of flux is the same as
that for the standard hierarchy $\nu=2/5$ state\cite{Haldane}.
This should not be a surprise, since some of the first trial
wavefunctions for $\nu=2/5$ (for fermions) in the hierarchy were
based on pairing\cite{Halperin}. In the appendix we analytically
establish that this state is the unique zero energy state of the
Hamiltonian $P^3_3$ at this flux. We have numerically confirmed
this fact  by explicitly diagonalizing the Hamiltonian $P^3_3$
with up to $N=12$ particles on the spherical geometry and up to
$N=10$ particles on the torus. (We note that on the torus the
Gaffnian occurs at $N_{\phi} = (3/2+q) N$ meaning there is no
``shift", which is always the case on the torus).

In appendix \ref{app:further} we consider possible generalizations
of the form of the Gaffnian wavefunction (Eq.~\ref{eq:gaffwf}). In
particular, we find trial states for wavefunctions of the Jain
series $\nu=p/(m p + 1)$ (with $m$ odd for bosons and even for
fermions) with the same value of the flux as the usual Jain
sequence.  Since (as we will discuss below) the Gaffnian is
distinct from the hierarchy (or Jain) states, we suspect that
these trial states are similarly distinct from the usual Jain
states. However, we leave detailed study of these wavefunctions
for further work.

Since the Gaffnian is a paired state\cite{Halperin,Moore}, we
expect that each additional flux added will correspond to two
quasiholes, each with charge $e^* = e \nu/2$  with $-e$ the charge
on the elementary underlying ``electron" (or underlying boson for
$q=0$).  Generally, we define the number of extra flux added to
the Gaffnian ground state to be
\begin{equation}
n = N_{\phi} - (3 N/2  - 3 +  q (N-1))
\end{equation}
(compare Eq. \ref{eq:flux}).   Note that $n$ here is defined so
that it is half integer if $N$ is odd.   To construct
wavefunctions in the presence of $n$ (integral) additional flux,
we can insert a factor of
\begin{equation}
\label{eq:insert}
 \prod_{a \leq N/2; j \leq n}(z_a - w_j)
\prod_{N/2 <\, b \leq N;  n < k \leq 2n} (z_b - w_k)
\end{equation} into the above wavefunction inside the
symmetrization where the $w$'s indicate the quasihole positions.
However, for $2n$ fixed quasihole positions, there are apparently
$2n \choose n$ inequivalent ways to choose which of the positions
$w_j$ are labelled with an index $j \leq n$ and which with an
index $j > n$. One might expect that the different groupings of
the positions into these two groups generate equally many
inequivalent quasihole wavefunctions.  The fact that we find more
than one independent quasihole wavefunction means that there are
zero-modes associated with these quasiholes\cite{Moore}. Analogous
to the Pfaffian\cite{ReadRezayiPfaff}, however, it turns out that
there are many linear dependencies between these many different
wavefunctions.

In appendix \ref{app:counting}, we show explicitly how to count
the zero energy ground state degeneracy of the system with
Hamiltonian $P^3_3$ at any flux (Strictly speaking the appendix
only addresses the case of $N$ and $2n$ even, although the odd
case proceeds similarly). We find that the degeneracy of zero
energy states is given by
\begin{equation}
    \label{eq:decompose}
    \sum_{F, (-1)^F = (-1)^N}^{F_{max}} {(N-F)/2 + 2n \choose 2n} {n + F/2 -1 \choose F}
\end{equation}
where the maximum value of $F$ is given by $F_{max} =
\min(N,2n-2)$.   To verify this result, we have numerically
performed exact diagonalizations. For every case we have examined,
we find perfect agreement between this analytic rule and the
results of our exact diagonalization of the Hamiltonian $P^3_3$.
(We have examined $N=4, 6$ with $n \leq 6$, $N=8$ with $n \leq 4$,
$N=10$ with $n \leq 2$, $N=5$ with $n \leq 7/2$ and $N=7$ with $n
\leq 5/2$.)

The first term in Eq.~\ref{eq:decompose} corresponds to the
positional degeneracy of the quasiholes and can be thought of as
$2n$ bosons in $(N-F)/2 + 1$ orbitals.  The second term is the
degeneracy of the zero-modes and can be thought of as $F$ fermions
in $n+F/2-1$ orbitals.  Since the number of orbitals changes half
as fast as number of particles, these zero-modes are a realization
of semionic exclusion statistics\cite{Exclusion}.

The form of Eq.~\ref{eq:decompose} is quite analogous to the
zero-mode counting expressions found for the
Pfaffian\cite{ReadRezayiPfaff}, Haffnian\cite{Green}, and
Read-Rezayi States\cite{ReadRezayi,Gurarie}.  However, in those
cases the zero-modes have fermionic, bosonic, and parafermionic
statistics respectively.  For the fermionic (Pfaffian) case we put
$F$ fermions in a fixed number $n$ orbitals\cite{ReadRezayiPfaff}.
For the bosonic (Haffnian) case\cite{Green}, we put $F$ fermions
in $n+F-2$ orbitals (which is equivalent to putting $F$ bosons in
$n-1$ orbitals).  The Gaffnian case is quite naturally an
interpolation between these two cases.  (The Read-Rezayi
parafermion case cannot be phrased in this language so
easily\cite{Gurarie}).

 As with the  Pfaffian\cite{ReadRezayiPfaff}, Haffnian\cite{Green}, and
Read-Rezayi\cite{ReadRezayi,Gurarie} cases, the structure of
Eq.~\ref{eq:decompose} also tells us how to decompose these
degenerate states into angular momentum multiplets.  We simply
calculate the multiplets of the $2n$ bosons in $(N-F)/2 + 1$
orbitals and also the multiplets of the $F$ fermions in $n+F/2-1$
orbitals and then add these together using standard angular
momentum addition rules.  An explicit example of this angular
momentum addition is given in appendix \ref{app:explicit}.

As discussed above, we can also look at wavefunctions with fixed
quasiparticle positions. The number of linearly independent states
should just be given by the zero-mode contribution to the above
equation  \begin{equation} \label{eq:fixedsum}
 D_n = \sum_{F, (-1)^F = (-1)^N}^{F_{max}} {n + F/2 -1 \choose F}
\end{equation}
Indeed by generating wavefunctions (described by Eq.
\ref{eq:insert} inserted into Eq. \ref{eq:gaffwf}) numerically and
checking for linear independence, we find precisely this number of
independent states for all cases we have tried ($N=4, n\leq 6$;
$N=6, n \leq 5$, $N=8, n \leq 3$).

It is interesting to note that in the case of $N \geq 2n -2$ (so
$F_{max} = 2n-2$) the sum  Eq. \ref{eq:fixedsum} gives the
$2n-1^{st}$ Fibonacci number, $\mbox{Fib}(2n-1)= \mbox{Fib}(N_{qh}
-1)$.  This can be proven trivially by induction on $n$ to show
that $D_n + D_{n+1/2} = D_{n+1}$.   We note that the $Z_3$
Read-Rezayi state where also has a degeneracy of
$\mbox{Fib}(N_{qh}-1)$.  Another similarity we have found is that
both states have a 2-fold degeneracy of the ground state at zero
momentum on the torus geometry (in addition to the usual center of
mass degeneracy\cite{Prange}).   However, the two ground states
occur at different values of the flux for a finite spherical
system, so they are topologically different states.  Also, as
mentioned above, the state counting formula analgous to
Eq.~\ref{eq:decompose} involves parafermionic\cite{Gurarie}
zero-modes for the Read-Rezayi case compared to semionic modes for
the Gaffnian.

\begin{figure}[tbph]
 \hspace*{10pt} \begin{minipage}{2.5in}
\begin{tabular}{|c||c|} \hline
$ $ & $\Delta$  \\
\hline \hline $\psi$ & 3/4  \\
 $\varphi$ & 1/5  \\
$\sigma$ & -1/20
\\ \hline
\end{tabular}
\hspace*{25pt}
\begin{tabular}{|c||c|c|c|} \hline
$\times$ & $\psi$ & $\varphi$ & $\sigma$ \\
\hline \hline $\psi$ & {\bf 1} &  &  \\
 $\varphi$ & $\sigma$ &  ${\bf 1} + \varphi$ &  \\
$\sigma$ & $\varphi$ & $\sigma +\psi$ & ${\bf 1} + \varphi$
\\ \hline
\end{tabular}
\end{minipage}
\vspace*{5pt}
 \caption{In the Virasoro minimal model conformal field theory
${\cal M}(3,5)$, there are three nontrivial fields, $\psi$,
$\varphi$ and $\sigma$ with dimensions $\Delta$ given in the left
table and fusion algebra given in right table. \label{fig:fusion}}
\end{figure}

\section{Conformal Field Theory}
\label{sec:conformal}

\begin{figure}[tbph]
\vspace*{5pt} \setlength{\unitlength}{1mm}
\begin{picture}(60,20)(-14,0)
\put(0,0){\vector(1,1){5}} \put(5,5){\vector(1,1){5}}
\put(5,5){\vector(1,-1){5}} \put(10,0){\vector(1,1){5}}
\put(10,10){\vector(1,1){5}} \put(10,10){\vector(1,-1){5}}
\put(15,5){\vector(1,1){5}} \put(15,5){\vector(1,-1){5}}
\put(15,15){\vector(1,-1){5}} \put(20,0){\vector(1,1){5}}
\put(20,10){\vector(1,1){5}} \put(20,10){\vector(1,-1){5}}
\put(25,5){\vector(1,1){5}} \put(25,5){\vector(1,-1){5}}
\put(25,15){\vector(1,-1){5}} \put(30,0){\vector(1,1){5}}
\put(30,10){\vector(1,1){5}} \put(30,10){\vector(1,-1){5}}
\put(35.5,0){\ldots} \put(35.5,10){\ldots} \put(0,-3){0}
\put(10,-3){2} \put(20,-3){4} \put(30,-3){6} \put(-5,0){$\bf{1}$}
\put(-5,5){$\sigma$} \put(-5,10){$\varphi$} \put(-5,15){$\psi $}
\end{picture}
\vspace*{5pt}
 \caption{The
Bratteli diagram shows how the $2n$ quasihole fields $\sigma$ fuse
together.  This is just a graphical representation of the fusion
rules (Table \ref{fig:fusion}) where at each horizontal step, the
states at the previous horizontal position are fused with one more
$\sigma$ field.   The number of conformal blocks
--- which gives the nonabelian degeneracy --- is seen graphically
by the number of paths through the diagram starting and ending at
the bottom when $N$  (and $2n$) is even.  When $N$ (and $2n$) is
odd, the path needs to start at the bottom but end at the top to
fuse with the one unpaired $\psi$ field.  By straightforward
counting, the number of such paths with $2n$ steps can be seen to
be the $2n-1^{st}$ Fibonacci number.} \label{fig:table}
\end{figure}
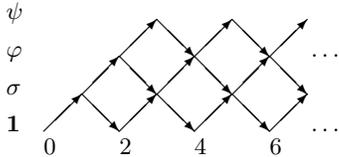

We now write this Gaffnian wavefunction as a correlator of a
conformal field theory (CFT)\cite{Conformal}.   Making the
connection to CFT has, in the past, been extremely powerful in
understanding states with nonabelian statistics (See for example
Refs.~\onlinecite{Moore,ReadRezayi,Gurarie,Slingerland}). For
example, the structure of a CFT can tell us about behavior of the
degenerate space under adiabatic braiding of
quasiholes\cite{Slingerland}. A CFT describing a paired state
should contain a field $\psi$ with fusion relation $\psi \times
\psi \sim {\bf 1}$ such that it has operator product expansion
\begin{equation}
    \psi(z) \psi(w) \sim (z-w)^{-2 \Delta_\psi}[{\bf 1} + \ldots ]
\end{equation}
 with $\bf 1$ the identity, $\Delta_\psi$ the conformal weight
(or dimension) of $\psi$, and dots representing less singular
terms. We can then construct a paired wavefunction
\begin{equation}
\Psi = \left\langle \prod_{i=1}^N \psi(z_i) \right\rangle
    \prod_{i < j} (z_i-z_j)^{2 \Delta_\psi + q}
\end{equation}
 Repeating the arguments which are presented in
Ref.~\onlinecite{ReadRezayi} it is clear that (for $q=0$, i.e.,
for bosons) this wavefunction will not vanish as 2 particles come
to the same position since the (fractional) Jastrow factor
precisely cancels the singularity of the operator product
expansion.  However, the wavefunction vanishes as $z^{4
\Delta_\psi}$ powers when the third particle approaches the other
two (since there are three (fractional) Jastrow factors and only
one singularity).  The Moore-Read Pfaffian \cite{Moore} is
described in this way by the Ising CFT, also known as the ${\cal
M}(4,3) $ minimal model\cite{Conformal}, which contains such a
field $\psi$ with weight $\Delta_\psi = 1/2$ so the wavefunction
vanishes as $z^2$ as three particles come to the same point.  The
Gaffnian is correspondingly described by one of the simplest
generalizations of the Ising CFT, known as the minimal model
${\cal M}(5,3)$. This CFT has a field $\psi$ with $\Delta_\psi =
3/4$ so that the wavefunction vanishes as $z^3$ as three particles
coalesce (for $q=0$). The dimensions and fusion rules for the
three independent fields ($\psi, \sigma$ and $\varphi$) in this
model are given in Fig.~\ref{fig:fusion}. The fusion of the field
$\sigma$ with the field $\psi$ gives us the operator product
expansion\cite{Conformal} \begin{equation} \psi(z) \sigma(w) \sim
(z-w)^{-1/2} \varphi(w) + \ldots  \label{eq:ope2}
\end{equation}
where here the exponent $-1/2$ is determined by the conformal
weights in Fig.~\ref{fig:table} as $ \Delta_\varphi -\Delta_\psi -
\Delta_\sigma$. As described in Ref.~\onlinecite{ReadRezayi} this
power of $1/2$ means that the quasihole created by the field
$\sigma$ must have charge $e^* = e \nu/2$ consistent with our
expectation for a paired state.  To see how this happens we write
a general wavefunction in the presence of $2n$ quasiholes as
\begin{eqnarray} \nonumber
\Psi &=& \left\langle \prod_{j=1}^{2n} \sigma(w_j) \prod_{i=1}^N
\psi(z_i) \right\rangle \\ & &
    \prod_{i < j} (z_i-z_j)^{3/2 + q} \prod_{i=1}^N
    \prod_{j=1}^{2n} (z_i - w_j)^{1/2} \label{eq:cwf2}
\end{eqnarray}
Given the operator product expansion Eq. \ref{eq:ope2}, the final
exponent in  Eq. \ref{eq:ope2} must have power 1/2 so that the
wavefunction is single valued in the $z$'s.   This Jastrow factor
then pushes precisely a charge $e \nu/2$ away from each quasihole.

We can also use the fusion rules to count the degeneracy of the
$2n$ quasihole state. The degeneracy is given by the number of
ways the $\sigma$ fields in Eq. \ref{eq:cwf2} can fuse together to
form the identity.   This is illustrated graphically as the number
of paths through the Bratteli diagram\cite{Slingerland} shown in
Fig.~\ref{fig:table}. The number of paths is $\mbox{Fib}(2n-1)$,
which is consistent with the result of our above counting formula.
If the number of particles $N$ is even, then we pair the $\psi$
fields to form identities, and the $\sigma$ fields must also pair
to form the identity. However, if $N$ is odd, we can only form the
identity if the $\sigma$ fields fuse to form one more $\psi$ that
can then fuse to form the identity with the one remaining $\psi$
field.

One may ask how we know that we have the correct conformal field
theory (particularly in light of the fact that classification of
all conformal field theories is an ongoing research field). The
fact that we have a paired state at filling fraction $\nu=2/3$ for
bosons (i.e., the fact that the wavefunction does not vanish when
two particles come together) means we must have a field $\psi$
which fuses with itself to form the identity.  The fact that the
Hamiltonian forces the wavefunction to vanish as three powers when
three particles come together further fixes the dimension
$\Delta_\psi$. It is easy enough to show that the only Virasoro
minimal model conformal field theory with such a field is ${\cal
M}(5,3)$. If we further insist that the charge of the quasihole
should be $e \nu /2$, as is expected for a paired state, this
fixes the exponent of the final factor in Eq. \ref{eq:cwf2}, and
this in turn fixes $\Delta_\varphi-\Delta_\sigma = 1/4$. We must
also insist that the fusion relations for fusing many
quasiparticles with each other has the form of the Bratteli
diagram in Fig.~\ref{fig:table}.   Finally, one can look at the
subleading behavior of the wavefunction as particles approach each
other to extract the central charge of the theory, which again is
consistent with ${\cal M}(5,3)$ (we do not perform this
calculation here).  These restrictions place serious constraint on
any possible conformal field theory we would like to use to
represent the gaffnian state.   Certainly there is no ``simple"
(i.e., minimal model) theory other than $M(5,3)$ with the required
properties. However, we have not proven that no other theory
exists.

The conformal field theory ${\cal M}(5,3)$ is
nonunitary\cite{Conformal}.  This highly suggests that the
Gaffnian wavefunction does not represent a true phase, but rather
represents a quantum critical point.  The argument for this goes
as follows:  The edge state theory in 1+1 dimensions of a quantum
Hall state should be described by the same conformal field theory
as the bulk 2 dimensional theory.   However, since the edge state
theory is a dynamical theory, it must be unitary.  If we have a
trial wavefunction that is generated by a nonunitary theory,
apparently the only way out of this conundrum is that the edge
state theory does not exist; i.e, edge excitations do not stay on
the edge, but leak into the bulk. This could indeed be the case if
the ground state has arbitrarily low energy excitations in the
thermodynamic limit. This could in turn occur if the wavefunction
represents a quantum critical point.  Indeed, there have been past
examples of critical quantum Hall states which are described by
nonunitary CFTs\cite{ReadRezayiPfaff,Green,GreenRead}.  While
there is no strict proof that a nonunitary conformal field theory
necessarily implies a critical state, there is also no
understanding of how anything else could occur.

\section{Exact Diagonalizations}
\label{sec:exact}

We now turn to exact diagonalizations.  Strictly speaking, the
Hamiltonian $P^3_3$ has been defined to be a projection operator
that acts on the full wavefunction (to keep any states where any
three particles have relative angular momentum less than three).
As such, this Hamiltonian has eigenvalues that are either zero
(for the zero energy space) or unity.   A more physical version of
this Hamiltonian can be written as
\begin{equation}
\label{eq:Hloc}  H = \sum_{i < j< k} \left( V_{3,0} P^0_{ijk} +
V_{3,2} P^2_{ijk}\right)
\end{equation}
where we have defined a general three body operator $P_{ijk}^p$
which projects out (i.e., keeps) any component of the wavefunction
where the three particles $i$, $j$ and $k$ have relative angular
momentum $L_{min} + p$.   (On the sphere\cite{Haldane,Us}, one
defines $P_{ijk}^p$ to project out (i.e., keep) any cluster of
three particles with total angular momentum $3 N_{\phi}/2 - p$).

\begin{figure}[htbp]
\scalebox{.3}{\includegraphics*{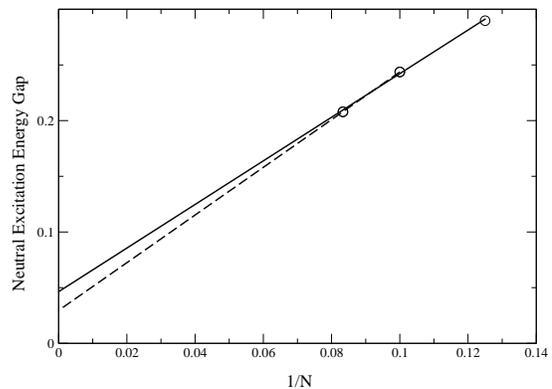}}  \caption{Lowest neutral gap
 excitation of the Gaffnian
as a function of system size (using the Hamiltonian in Eq.
\ref{eq:Hloc}) in units of $V_{3,0} = V_{3,2}$.  Data is shown for
$N=8,10,12$ particles.  The solid is a linear fit of all three
data points. The dashed line is a fit of the two larger systems
only (suggesting that if we could access even larger systems, the
extrapolations might be even closer to zero).   This data suggests
the possibility that the gap may extrapolate to zero in the
thermodynamic limit, as would be expected for a critical state.
However, from the available numerical data, we cannot exclude the
possibility that it extrapolates to a finite value.}
\label{fig:gaps}
\end{figure}

Note that three particles cannot\cite{Us} have relative angular
momentum of $L_{min} + 1$, so this Hamiltonian gives energy to any
case where the relative angular momentum of any cluster of three
particles is less than three.   ( Some readers may have assumed
that the form of Eq. \ref{eq:Hloc} is what we meant all along when
we have been writing $P^3_3$, as we were not very explicit about
what we meant.)   Since the Hamiltonian Eq. \ref{eq:Hloc} gives
energy to any cluster of three particles with relative angular
momentum less than 3, it has precisely the same zero energy space
as $P^3_3$. However, the excitation spectrum here is different,
and is dependent on the values of $V_{3,0}$ and $V_{3,2}$.

Let us first examine the issue of criticality.  In
Fig.~\ref{fig:gaps}, we show the lowest energy neutral excitation
of $H$   (from Eq. \ref{eq:Hloc}) as a function of system size for
$N=8,10,12$ on a spherical geometry with $V_{3,0} = V_{3,2}$ (We
have chosen to look at bosons on a sphere because we can go to
larger systems). As can be seen in the figure it appears that the
gap extrapolates to a positive value, but it is not possible to
rule out extrapolation to a zero value which would be a sign of
criticality. Furthermore, changing the ratio of $V_{3,0}/V_{3,2}$
(data not shown) does not appear to substantially affect the ratio
of the extrapolated energy to the reference energy of the gap for
$N=10$.

\begin{figure}[htbp]
\scalebox{.35}{\includegraphics*{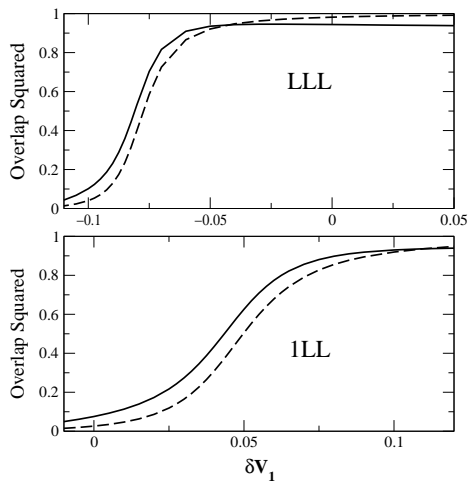}} \caption{ Squared
overlaps of  trial states with the exact ground state at $\nu=2/5$
on a sphere with $10$ electrons, as we vary the coulomb
interaction.  Solid line is the overlap of the Gaffnian
wavefunction with the exact ground state. The dashed line is the
hierarchy $2/5$ state\cite{endnote4} with the exact ground state.
The top is results for the lowest Landau level, the bottom is the
first excited Landau level.   In the horizontal direction the
interaction is varied around the Coulomb interaction by adding an
additional $\delta V_1$ Haldane pseudopotential. }
\end{figure}

We now turn to the question of whether the Gaffnian is physically
relevant to the physics of 2D electron systems.  We have performed
exact diagonalization on a spherical geometry for $10$ electrons in
the lowest Landau level (LLL) and first excited Landau Level (1LL),
and we have varied the electron-electron interaction in the
neighborhood of the Coulomb interaction by varying the Haldane
pseudopotential\cite{Prange,Haldane} coefficient $V_1$. In Fig.~2a,
we show the overlap of the exact ground state with our trial
wavefunction.  Results are shown for the Gaffnian (solid) and the
hierarchy 2/5 state\cite{Prange,endnote4} (dashed).  Over a range of
$V_1$ both trial states have quite good overlaps with the ground
state considering that the zero angular momentum Hilbert space has
52 dimensions  (Note that for many of the well known numerical
cases\cite{Olle,Prange} where extremely large overlaps have been
reported, the dimension of the available Hilbert space is much
smaller than this).   Near the regime of $V_1$ where the overlaps
drop, we believe the system is in the Read-Rezayi
phase\cite{EdandNick} (although at a different value of flux on the
sphere).   Since both the Gaffnian and hierarchy states have such
large overlaps with the ground state, they necessarily have large
overlaps with each other, although in the thermodynamic limit they
become orthogonal.

We have also performed exact diagonalization on the torus
geometry.   Here, the Gaffnian ground state is found to be doubly
degenerate (in addition to the usual center of mass degeneracy).
The two zero energy ground states are distinguished by a parity
quantum number.   The state with positive parity again has
extremely high overlap with the hierarchy state, similar to the
overlaps on the sphere.  As on the sphere, both of these have a
high overlap with the exact ground state for a wide range of
interactions.   However, we do not find that the exact ground
state has an even approximate double degeneracy in the regimes
where the overlaps of the Gaffnian and the hierarchy are large.
Approximate double degeneracy of the ground state is found where
we believe the Read-Rezayi state is the proper ground
state\cite{EdandNick}.

\section{Discussion}
\label{sec:discussion}

If the Gaffnian does turn out to be a critical state, as suggested
here, this then raises the question as to what the neighboring
phases are.   It is reasonable that one would be a ``strong
pairing" phase (albeit one that cannot be easily described within
BCS theory\cite{GreenRead}) which may correspond to the hierarchy
wavefunction itself\cite{Halperin}.  This would be quite natural
considering the high mutual overlaps of the hierarchy and
Gaffnian.

The nature of the state on the opposite side of the transition is
a bit harder to guess at.  One possibility is that it is the
Read-Rezayi state.  This would make some sense because of the
similar ground state degeneracy.  Here, we imagine that as we
approach the transition from the hierarchy side, the putative zero
energy state would drop continuously and hit zero energy at the
Gaffnian critical point. It would then stay at zero energy through
the Read-Rezayi phase.    On the other hand, we should note that
there is a notable  topological difference between the Read-Rezayi
and Gaffnian state, which is more evident on the sphere as they
occur at different values of flux.

Yet another possible candidate for a state that might occur nearby
is a charge density wave state.  We leave the project of sorting
out the details of this transition for future work.

If the Gaffnian is in fact a critical state, this means that the
concept of ``nonabelian" statistics\cite{Moore} may not be well
defined. Indeed, the idea of statistics describes what happens to
a system when particles are adiabatically exchanged. Since
definition of adiabatic usually requires any perturbation to the
system to be on a timescale slower than $\hbar/\Delta$ with
$\Delta$ the minimum gap in the system, if the system has gapless
excitations, there is generally no way to have adiabaticity.  One
might ask whether any remnant of the idea of nonabelian statistics
still remain.  This is a question that is hard to answer without
knowing the details of what these ``critical" low energy
excitations are.

{\bf Acknowledgements:}  This work was partially supported by the
UK EPSRC Grant No. GR/R99027/01 (NRC) and by the US DOE contract
DE-FG03-02ER45981 (EHR).  Conversations with F. D. M. Haldane, N.
Read, and V. Gurarie are gratefully acknowledged.  The authors
particularly thank N. Read for giving feedback on an earlier draft
of this work, and for emphasizing the likely criticality of a
nonunitary theory.

\appendix

\section{Analytic Counting of Zero Energy States}
\label{app:counting}

In this appendix we will enumerate all possible zero energy states
of the Hamiltonian $P^3_3$ on the sphere with any number of
particles and at any given flux.   Our approach will be in two
parts.  In section \ref{sec:degen} we will write down a linearly
independent set of zero energy states (and we will count them). It
is this section that shows most clearly how the semionic
zero-modes arise. Then in section \ref{sec:complete1} we will show
that these wavefunctions are indeed zero energy states of $P^3_3$.
Finally, we will show in \ref{sec:complete2} that these states
indeed form a complete set of the zero energy states.   The
arguments here are quite similar to those given in Refs.
\onlinecite{ReadRezayiPfaff,Green}. However, here the situation is
more complicated as our zero-modes are not simple fermions and
bosons as in those two references.  Note that throughout this
appendix we will focus on the case where $N$ (and therefore $2n$)
is even. The $N$ odd case is a relatively simple generalization.

\subsection{Counting States} \label{sec:degen}


We start with the requirement that $\psi$ vanishes as three powers
as any three particles approach each other.   The wavefunction Eq.
\ref{eq:gaffwf} clearly provides one such solution (at a given
value of flux).  We will call this the Gaffnian ground state.  In
this section we propose a more general form for wavefunctions when
there are some arbitrary number $2n$ quasiholes (or $n$ additional
flux) added to the ground state, and we will count the number of
such states that are linearly independent.

Inspired by the work of Refs. \onlinecite{ReadRezayiPfaff,Green}
and analogous to the Pfaffian, Haffnian and Haldane-Rezayi states,
we write our proposed wavefunction in a form with broken and
unbroken pairs. Let us declare that $F$ of the $N$ total particles
are unpaired. Restrictions on $F$ will be determined later. We
then propose the following form for our wavefunctions:
\begin{widetext}
\begin{eqnarray}\nonumber
\psi_G &=& \tilde S \left[ \prod_{1 \leq a < b \leq \frac{N}{2}}
(z_a - z_b)^{2+q} \prod_{\frac{N}{2}+1 \leq c<d \leq N} (z_c -
z_d)^{2+q} \prod_{1 \leq e \leq \frac{N}{2} < f \leq N} (z_e -
z_f)^{1+q} \right. \\ & &  \left. \prod_{1 \leq g \leq
\frac{N-F}{2}}
\frac{\Phi(z_{g+\frac{F}{2}},z_{g+\frac{N+F}{2}};w_1,\ldots,w_{2n})}{z_{g+\frac{F}{2}}
- z_{g+\frac{N+F}{2}}} \left(\prod_{i=1}^{\frac{F}{2}}
\prod_{j=\frac{N}{2}+1}^{\frac{N+F}{2}} (z_i - z_j)^{1+q}
\;\Omega
(z_1,\ldots,z_{\frac{F}{2}};z_{\frac{N}{2}+1},\ldots,z_{\frac{N+F}{2}})
\right) \right]
\end{eqnarray}
\end{widetext}
where as above $\tilde S$ either symmetrizes (for bosons, even
$q$) or antisymmetrizes (for fermions, odd $q$) over all of the
$z$ coordinates.  Here we have defined $\Phi$ to be the
Read-Rezayi quasihole insertion\cite{ReadRezayiPfaff}:
\begin{eqnarray}
\label{eq:RRpairingphi}
 & & \Phi(z_1,z_2;w_1,\ldots,w_{2n}) =  \\ & &
\frac{1}{(n!)^2} \sum_{\tau \in S_{2n}} \prod_{i=1}^{n} (z_1 -
w_{\tau(2i-1)})(z_2 - w_{\tau(2i)}) \nonumber
\end{eqnarray}
and $\Omega$ is a wavefunction for the zero-modes to be
determined later.  (The sum over $\tau \in S_{2n}$ is the sum
over permutations of the $2n$ variables $w$).  We now specialize
to the case of $q=0$ (bosons) for simplicity.   Since the particles must be
at the flux $N_\phi = 3(\frac{N}{2}-1)+n$ we will deduce that the
highest degree of the unpaired particle coordinates is $n-1$.  To
see how this is deduced we start by defining $A \equiv \{z_1,
\ldots, z_{\frac{N}{2}}\}$, and $B \equiv \{z_{\frac{N}{2}+1},
\ldots, z_N\}$).   We then simply count up powers of $z_k$
appearing in $\psi_G$
\begin{widetext}
\begin{equation}
\psi_G = S \Bigl[\underbrace{ \underbrace{\prod_{i<j} (z_{A_i} -
z_{A_j})^2}_{2(\frac{N}{2}-1)\text{ for A}}
\underbrace{\prod_{i<j} (z_{B_i} -
z_{B_j})^2}_{2(\frac{N}{2}-1)\text{ for B}}
\underbrace{\prod_{i,j} (z_{A_i} - z_{B_j})}_{\frac{N}{2} \text{
for A\&B}} }_{\text{both paired \& unpaired}}
\underbrace{\underbrace{\prod_i
\rule[-15pt]{0pt}{15pt}\frac{\Phi(z_{A_i}, z_{B_i})}{(z_{A_i} -
z_{B_i})}}_{-1+n  \rule[-2pt]{0pt}{2pt} \text{ for
A\&B}}}_{\text{paired only}} \underbrace{\underbrace{\prod_{i,j}
(z_{A_i} - z_{B_j})}_{\frac{F}{2} \text{ for
A\&B}}\underbrace{\rule[-15pt]{0pt}{15pt}\Omega(\ldots)}_{???
\text{ for A\&B}}}_{\text{unpaired only}}\Bigr]
\end{equation}
\end{widetext}
Here $S$ symmetrizes over all coordinates $z$.
In this equation,  we have written beneath each term the number of
powers of $z_k$ occurring.  Thus adding up the powers, we conclude
that $\Omega$ is some polynomial in unpaired coordinates of
degree $m_i: 0 \leq m_i \leq n-1-\frac{F}{2}$ for each unpaired
coordinate $z_i$. Notice also, that this puts a restriction on
$F$: $F \leq 2n-2$, and since obviously $F \leq N$, we obtain
\begin{equation} F \leq
  \min(2n-2, N) \end{equation}
  as written above in the main text.  The maximum
  degree of $\Omega$ occurs when $F=0$ and is given by
  $n-1$.


To see how many linearly independent wavefunctions we have for
given $\{N, n, F\}$ we proceed as follows.  We choose $N-F$
(necessarily even here) of the $N$ coordinates and group them
together in pairs  $\{(z_{a_i},
  z_{b_i})\}$ for $i=1, \ldots (N-F)/2$ with $a_i, b_i \in [1 \ldots N]$ and $a_i \neq a_j$, $b_i \neq b_j$ for
  $i \neq j$ and $a_i \neq b_j$ for all $i,j$.
  We then bring together the position of the
paired paired particles to coordinates $\tilde z_i$.  In other
words we set $z_{a_i} = z_{b_i} = \tilde z_i$ for $i = 1, \ldots
(N-F)/2$.  Taking this limit selects out a particular group of
terms from the original full symmetrization that do not vanish. In
particular the nonvanishing terms are the terms in which a factor
of  $\Phi(z_{a_i}, z_{b_i};\ldots) (z_{a_i} - z_{b_i})^{-1}$
appeared
  for each pair $(z_{a_i}, z_{b_i})$. The other terms will have a
  factor of $(z_{a_i} - z_{b_i})$ in the numerator, and will
  vanish in these limits.  After taking these limits we are left with
\begin{widetext}
  \begin{eqnarray}
    \tilde{\psi} = S^{\prime}\left[
    \underline{\prod_{i<j} (\tilde{z}_i - \tilde{z}_j)^6
    \prod_k \Phi(\tilde{z}_k;w_1, \ldots, w_{2n})}
    \prod_{1 \leq a < b \leq \frac{F}{2}} (z_a - z_b)^2
    \prod_{\frac{N}{2}+1 \leq c < d \leq \frac{N+F}{2}} (z_c - z_d)^2\right. \\ \left.
    \underline{\underline{\prod_{l=1}^{\frac{N-F}{2}} \prod_{e=1}^{\frac{F}{2}} (z_e - \tilde{z}_l)^3
    \prod_{m=1}^{\frac{N-F}{2}} \prod_{f=\frac{N}{2}+1}^{\frac{N+F}{2}} (\tilde{z}_m - z_f)^3 }}
    \prod_{g=1}^{\frac{F}{2}} \prod_{h=\frac{N}{2}+1}^{\frac{N+F}{2}} (z_g - z_h)^2 \,
    \Omega (z_1,\ldots,z_{\frac{F}{2}};z_{\frac{N}{2}+1},\ldots,z_{\frac{N+F}{2}})
    \right] \nonumber
  \end{eqnarray}
  \end{widetext}
  where $S^\prime$ symmetrizes over $ \{\tilde{z}_i\}$ and $\{z_1,
  \ldots, z_{\frac{F}{2}}, z_{\frac{N+1}{2}}, \ldots,
  z_{\frac{N+F}{2}} \}$ separately. In other words, $S^\prime$ is what remains of the
  original full symmetrization over the $N$ particles.   The underlined
  factors contain the dependence of $\tilde{\psi}$ on $\tilde{z}$, and
  are symmetric in $\{\tilde{z}_i\}$, while the doubly underlined
  factor is symmetric in $\{ z_1, \ldots, z_{\frac{F}{2}},
  z_{\frac{N+1}{2}}, \ldots, z_{\frac{N+F}{2}} \}$ as well. Thus, the
  symmetrization $S^\prime$ reduces to $S''$ which symmetrizes over unpaired particles only
  (because the expression is already symmetric in $\tilde{z}_i$), and we can rewrite the
  wavefunction as
\begin{widetext}
  \begin{equation}
    \tilde{\psi} =  (\widetilde \pLJ)^2 \, \, S'' \left\{
    \left[ \Omega(\ldots)
    \prod_{l=1}^{\frac{N-F}{2}} \prod_{e=1}^{\frac{F}{2}} (z_e - \tilde{z}_l)^3
    \prod_{m=1}^{\frac{N-F}{2}} \prod_{f=\frac{N}{2}+1}^{\frac{N+F}{2}} (\tilde{z}_m - z_f)^3
    \right] \times
    \left[
    \prod_{i<j} (\tilde{z}_i - \tilde{z}_j)^6
    \prod_k \Phi(\tilde{z}_k)
    \right]
    \right\}
  \end{equation}
  \end{widetext}
  where $\widetilde \pLJ$ is a Laughlin-Jastrow factor in the unpaired particle
  coordinates. We thus discover that $\Omega(\ldots)$ can always
  be taken to be fully symmetric in its arguments  (any nonsymmetric parts vanish when symmetrized).
  We can thus think of this as a bosonic wavefunction for the zero-modes.  However, we've already
determined the maximal degree of $\Omega(\ldots)$ to be
$n-1-\frac{F}{2}$. The minimal degree   is obviously $0$, so we
have a total of $n-\frac{F}{2}$ orbitals in  which to put $F$
bosons.
  There are
  $\binom{(n-\frac{F}{2})+F-1}{F} = \binom{n+\frac{F}{2}-1}{F}$ such
  linear independent wavefunctions.   This is equivalent to
  placing $F$ fermions in $n+F/2 - 1$ orbitals.  Since the number
  of orbitals changes half as fast as the number of particles we
  put in them, these particles have semionic exclusion
  statistics\cite{Exclusion}.


\subsection{Zero Energy} \label{sec:complete1}

We will continue on to demonstrate that the linearly independent
set of wavefunctions we have just written down is in fact a
complete set of zero energy states of the Hamiltonian $P_3^3$.
First, however, we show that these wavefunctions are indeed zero
energy states.   The wavefunction for any zero energy state must
vanish as three or more powers when three particle positions come
to the same point. On the sphere\cite{Haldane,Us}, this is
equivalent to restricting the total angular momentum of the
cluster of three particles to be no greater than $3 N_{\phi}/2 -
3$.

First we'll show that the proposed wavefunctions $\psi_G$ are zero
energy eigenstates of this Hamiltonian.  For the ground state,
i.e.  no additional flux ($n=0$) we have $N_\phi = 3(\frac{N}{2} -
1)$.  Let us look at the $(ijk) \equiv (z_i, z_j, z_k)$ triplet.
We want to know what the highest total angular momentum is for
this triplet in our wavefunction $\psi_G$. The wavefunction can be
rewritten (\`a la Haldane\cite{Haldane}) as a sum of terms
proportional to $ f_{\text{rel}}(z_i, z_j, z_k)
f_{\text{tot}}(z_i, z_j, z_k)$ where $f_{\text{rel}}()$ is an
eigenstate of $l_{ijk}$, the 3 particle relative angular momentum
operator, and $f_{\text{tot}}()$ is an eigenstate of $L_{ijk}$,
the 3 particle total angular momentum operator.  Note that, as
above, we will always focus on $q=0$ for simplicity (The $q \neq
0$ case is a relatively minor generalization). To find the total
angular momentum, we look at the maximal degree of $z_i^\alpha
z_j^\beta z_k^\gamma$ in $f_{\text{tot}}()$, and find the total
angular momentum $L = \max(\frac{1}{2}(\alpha+\beta+\gamma))$.  To
find the maximum total angular momentum we must consider all
possible ways to have chosen the triplet $(z_i,z_j,z_k)$ from the
many terms in the wavefunction.  In particular, we must look at
all cases of which coordinate is one of the paired variables, and
which is unpaired, as well as looking at which variable is an
$A$-coordinate, and which is a $B$ coordinate. Here we are looking
at the relative angular momentum of a given triplet in each of the
many terms of the symmetrization sum.  All possibilities are
enumerated next.

\vspace*{10pt}

$\bullet$ Case 1: $i,j,k \in A$, $i<j<k$.

Here we have
\begin{eqnarray} \nonumber
      \alpha &=& \deg_{z_i}\psi_G = 2[(\frac{N}{2} - 1) - 2] + \frac{N}{2} -
      1\\
      \beta &=& \deg_{z_j}\psi_G = 2[(\frac{N}{2} - 2) - 1] + \frac{N}{2} -
      1\\
      \gamma &=& \deg_{z_k}\psi_G = 2[\frac{N}{2} - 3] + \frac{N}{2} -
      1\nonumber
  \end{eqnarray}
Using $L = \frac{1}{2}(\alpha+\beta+\gamma)$ and with $N_{\phi} =
3(N/2  - 1)$ we obtain in this case $    L = \frac{3}{2}N_\phi -
6$.

$\bullet$ Case 2a:  $i,j \in A,\; i<j$; $k \in B$  with pairing of
the form $(ia)(jb)(ck)$, i.e. terms of the form
\begin{equation}
\frac{\Phi(z_i,z_a)}{(z_i - z_a)} \frac{\Phi(z_j,z_b)}{(z_j -
z_b)} \frac{\Phi(z_c,z_k)}{(z_c - z_k)}
\end{equation}
Here we have
    \begin{eqnarray}\nonumber
          \alpha &=& \deg_{z_i}\psi_G = 2[(\frac{N}{2} - 1) - 1] + (\frac{N}{2} - 1) - 1
          \\
          \beta &=& \deg_{z_j}\psi_G = 2[\frac{N}{2} - 2] + (\frac{N}{2} - 1) - 1
          \\
\nonumber          \gamma &=& \deg_{z_k}\psi_G = 2[\frac{N}{2} -
1] + (\frac{N}{2} - 2) - 1
\end{eqnarray}
Similarly, adding up these powers we obtain an angular momentum $
 L = \frac{3}{2}N_\phi - 4.
 $

$\bullet$ Case 2b: $i,j \in A,\; i<j$; $k \in B$ with pairing of
the form $(ia)(jk)$, i.e, terms of the form
\begin{equation}
\frac{\Phi(z_i,z_a)}{(z_i - z_a)} \frac{\Phi(z_j,z_k)}{(z_j -
z_k)}
\end{equation}
Here we have
    \begin{eqnarray}
    \nonumber
          \alpha &=& \deg_{z_i}\psi_G = 2[(\frac{N}{2} - 1) - 1] + (\frac{N}{2} - 1) -
          1\\
          \beta &=& \deg_{z_j}\psi_G = 2[\frac{N}{2} - 2] + (\frac{N}{2} - 1)
          \\
          \gamma &=& \deg_{z_k}\psi_G = 2[\frac{N}{2} - 1] + (\frac{N}{2} -
          2)\nonumber
    \end{eqnarray}
which results in an angular momentum $L = \frac{3}{2}N_\phi - 3$.

These cases are the only possibilities.  Thus the highest total
angular momentum for any triplet is $\frac{3}{2}N_\phi - 3$ and so
the proposed wavefunction $\psi_G$ is a zero energy eigenstate of
the Gaffnian Hamiltonian $P^3_3$  as claimed.

\subsection{Completeness} \label{sec:complete2}

Now we show that the proposed wavefunctions span the complete set
of zero energy states of the Gaffnian Hamiltonian.  To do this we
will construct the most general zero energy eigenstate and show
that it takes the form of our proposed wavefunction. Take the
following zero energy wavefunction $\psi_G = \pLJ^2 \phi_G$ where
here $\pLJ$ is the Laughlin-Jastrow factor for all of the
particles.  Consider the behavior of $\psi_G$ as particles in an
arbitrary triplet $(ijk)$ approach each other, while the other
particles remain far away from the three.
\begin{widetext}
\begin{equation}
\psi_G \propto \underbrace{(z_i - z_j)^2(z_j - z_k)^2(z_k -
  z_i)^2}_{\alpha+\beta+\gamma=6, \text{ part of } \pLJ^2}\underbrace{(z_i -
  z_j)^{q_{ij}}(z_j - z_k)^{q_{jk}}(z_k -
  z_i)^{q_{ki}}}_{\alpha+\beta+\gamma=q_{ij}+q_{jk}+q_{ki}=Q, \text{ part of } \phi_G}
\end{equation}
\end{widetext}
The wavefunction vanishes as $6+Q$ powers as these three particles
come together.  This is equivalent\cite{Haldane,Us} to saying that
the total angular momentum of three particles is $L =
\frac{3}{2}N_\phi - (6+Q)$  (Since we're on the sphere, the
maximum angular momentum of each particle is $\frac{N_\phi}{2}$.
Any relative angular momentum reduces the total by a corresponding
amount\cite{Haldane,Us}).  Furthermore, by analyticity of $\psi_G$
we must have $q_{mn} \geq -2$.

Now, in order for $\psi_G$ to be a zero energy state of the
Gaffnian Hamiltonian, we must have $Q \geq -3$ (so that the
relative angular momentum of the cluster is greater than or equal
to $3 = 6 + Q$).
 From here on we'll concentrate on the $\phi_G$ factor of the
eigenstates, restoring the ubiquitous $\pLJ^2$ at the end. Allowed
forms in the Laurent expansion of $\phi_G$ as $(ijk)$ approach
each other are
\begin{eqnarray}
\label{eq:onlyform}
&&  \frac{1}{(z_i - z_j)^2 (z_j - z_k)}  \\
\label{eq:secondone}
&& \frac{z_k - z_i}{(z_i - z_j)^2 (z_j - z_k)^2} \\
&& \frac{1}{(z_i - z_j) (z_j - z_k) (z_k - z_i)}
\label{eq:thirdone}
\end{eqnarray}
as well as the same terms with $(ijk)$ permuted.   However, it is
easy to see that the second two forms reduce to the the first since
expression \ref{eq:secondone} is equivalent to
\begin{equation}
\frac{-1}{(z_i -
  z_j)^2 (z_j - z_k)} + \frac{-1}{(z_i - z_j) (z_j -
  z_k)^2}
\end{equation}
and expression \ref{eq:thirdone} is equivalent to
  \begin{equation}
\frac{-1}{(z_k -
  z_i)^2 (z_j - z_k)} + \frac{-1}{(z_i - z_j) (z_k - z_i)^2}
\end{equation}
It follows then, that it's enough to consider forms of type of Eq.
\ref{eq:onlyform} for triplets $(ijk)$ (as well as the same form
with permutations of $(ijk)$).  When $(ijk) \rightarrow
\tilde{z}$, the most general zero energy eigenstate should have
the form
\begin{equation} \phi_G \propto \frac{F(z_i, z_j, z_k)}{(z_i -
z_j)^2 (z_j - z_k)}
\end{equation}
(or a form like this with any permutation of $(ijk)$) where
$F(\ldots)$ must be analytic (i.e., with no poles).

Now arbitrarily pair up and relabel the particles, e.g. $(z_{A_1},
z_{B_1})$, $(z_{A_2}, z_{B_2})$, $\ldots$, $(z_{A_{{N}/{2}}},
z_{B_{{N}/{2}}})$. Look at the most singular part of $\phi_G$ as
particles within these pairs approach each other, while pairs are
kept separated.
\begin{eqnarray} \nonumber
\phi_G &\propto& \frac{1}{(z_{A_1} - z_{B_1})^2} \frac{1}{(z_{A_2}
- z_{B_2})^2} \ldots \\
& & \ldots  \frac{1}{(z_{A_\frac{N}{2}} - z_{B_\frac{N}{2}})^2} \,
\,  \phi_{\frac{N}{2}}
\end{eqnarray}
Since we've isolated the most singular part of $\phi_G$, it's
clear that $\phi_\frac{N}{2}$ cannot contain factors $(z_{A_i} -
z_{B_i})^{-1}$. If we now consider triplets $(A_1,B_1,k)\;\forall
k \notin \{A_1, B_1\}$, and bring particle $k$ close to $(A_1,
B_1)$ it's clear that $\phi_\frac{N}{2}$ must contain a factor of
$(z_{A_1} - z_k)^{-1}$ or $(z_{B_1} - z_k)^{-1}$, but \textbf{not}
both, in order to satisfy the requirements on the pole structure
deduced above. We might be led to naively define
\begin{equation}
\label{eq:sumj} \phi_{\frac{N}{2}} = \sum_j \left( \prod_{m, n}
  \frac{\phi_{1,j}}{(z_{A_1} - z_{\mathcal{D}^j_m})(z_{\mathcal{C}^j_n} - z_{B_1})}\right)
\end{equation}
where $\mathcal{C}^j \cup \mathcal{D}^j = \{z_i\}; \:
\mathcal{C}^j \cap \mathcal{D}^j = \emptyset$, (i.e. a partition
of the set of particle coordinates), and $j$ indexes all possible
partitions.
However, the pole structure places further restrictions on the
sets $\mathcal{C}$ and $\mathcal{D}$. In particular, $A_i$ and
$B_i$ cannot both be in $\mathcal{C}$ or in $\mathcal{D}$.
Otherwise, supposing $A_i,B_i \in \mathcal{C}$, we have
(schematically) the following:
\begin{eqnarray} \nonumber
\phi_G &\propto& \frac{1}{(z_{A_1} - z_{B_1})^2} \ldots
\frac{1}{(z_{A_i} - z_{B_i})^2} \ldots \\ &\ldots& \left[\ldots
\frac{1}{(z_{A_1} - z_{A_i})(z_{A_1}
    - z_{B_i})} \ldots \right]
\end{eqnarray}
and we immediately recognize, that the triplet $(A_1,A_i,B_i)$ has
too many poles ($Q < -3$). We conclude, that for $i$-th pair only
one factor is allowed in $\phi_\frac{N}{2}$, either $(z_{A_1} -
z_{A_i})^{-1}$, or $(z_{A_1} - z_{B_i})^{-1}$. That is, the
partitions are such that $\mathcal{C}$ includes only one
representative of any pair, and $\mathcal{D}$ includes the
complementary member of this pair, i.e. necessarily $A_i \in
\mathcal{C}$, $B_i \in \mathcal{D}$ \textbf{or} $B_i \in
\mathcal{C}$, $A_i \in \mathcal{D}$.   At this point we can
recognize that our notation of $\mathcal{C}$ and $\mathcal{D}$ is
redundant, and that we can rewrite
\begin{equation}
\phi_{\frac{N}{2}} \propto \sum_{k\in\text{Partitions}} \left(
 \prod_{i \neq j} \frac{\tilde{\phi}_k}{(z_{A^k_i} - z_{B^k_j})} \right)
\end{equation}
where $\tilde{\phi}_k$ cannot contain any more poles, and the
conventions are that for all $k$ the same coordinates are paired
up, i.e.  $\{ A^k_i, B^k_i \} = \{ A^{k^\prime}_i, B^{k^\prime}_i
\}$ are equal as sets. The difference between partition $k$ and
partition $k^\prime$ is the order of coordinates within a pair,
i.e. we could have $z_{A^k_i} = z_{B^{k^\prime}_i},\; z_{B^k_i} =
z_{A^{k^\prime}_i}$, \textbf{or} $z_{A^k_i} =
z_{A^{k^\prime}_i},\; z_{B^k_i} = z_{B^{k^\prime}_i}$. Clearly the
sum over $k$ is a subset of symmetrization over all particles.

Finally, we should also include the exchange of pairs
$(z_{A^k_i},\; z_{B^k_i}) \leftrightarrow (z_{A^k_j},\;
z_{B^k_j})$, since the most singular part is symmetric under this
exchange, and arrive at
\begin{widetext}
\begin{equation}
\phi_{\frac{N}{2}} = \sum_{\text{pairings}}\;\;
\sum_{k\in\text{Partitions}} \left(\prod_{i
    \neq j} \frac{\tilde{\phi}_{k}}{(z_{A^k_i} - z_{B^k_j})} \right)
\end{equation}
Then for the particular choice of pairs we'll have
\begin{equation}
\phi_G \propto \prod_i \frac{1}{(z_{A_i} - z_{B_i})^2} \left[
 \sum_{\text{pairings}}\;\;\sum_{k\in\text{Partitions}} \left( \prod_{i \neq j}  \frac{\tilde{\phi}_{k}}{(z_{A^k_i} - z_{B^k_j})} \right)
\right]
\end{equation}
and recover the whole eigenfunction by symmetrization over all
particles and multiplication by the Jastrow factor squared:
\begin{equation}
\psi_G = \pLJ^2\; S \left\{ \prod_i \frac{1}{(z_{A_i} -
z_{B_i})^2} \left[
 \sum_{\text{pairings}}\;\; \sum_{k\in\text{Partitions}} \left( \prod_{i \neq j}  \frac{\tilde{\phi}_{k}}{(z_{A^k_i} - z_{B^k_j})} \right)
\right]\right\}
\end{equation}

The wavefunction of the densest state has as few zeros as possible,
and to find it we may choose $\tilde{\phi}_k \equiv 1$, then
\begin{eqnarray}
\psi_G &=& \pLJ^2\; S \left\{
  \sum_{\text{pairings}}\;\; \sum_{k\in\text{Partitions}} \left(  \prod_i \frac{1}{(z_{A^k_i} - z_{B^k_i})^2}
  \prod_{i \neq j}  \frac{1}{(z_{A^k_i} - z_{B^k_j})} \right)
\right\} \equiv \\
& & \pLJ^2\; S\left[
 \prod_i \frac{1}{(z_{A_i} - z_{B_i})^2} \prod_{i \neq j}
 \frac{1}{(z_{A_i} - z_{B_j})}
\right]
\end{eqnarray}
\end{widetext}
This is can be recognized as the proposed Gaffnian wavefunction
with no broken pairs and no added flux.
To obtain the states of lower density we need to consider the case
of non-constant $\tilde{\phi}_k(z_{A_1}, z_{B_1}; \ldots;
z_{A_{N/2}}, z_{B_{N/2}})$. By analyzing the symmetry of
denominators we find that $\tilde{\phi}_k$ must be symmetric under
the exchange of pairs $(z_{A_i}, z_{B_i}) \leftrightarrow
(z_{A_j}, z_{B_j})$.  We now claim that a complete basis for
functions that satisfy this symmetry condition is given by
functions of the form
\begin{equation} \sum_{\tau \in
S_\frac{N}{2}} \prod_{i=1}^{N/2} f_{i}(z_{A_{\tau(i)}},
z_{B_{\tau(i)}})
\end{equation}
where the $f_i$'s are chosen from a basis for arbitrary
polynomials of their two arguments.    While this may seem to be a
strange way to write a basis for the polynomial
$\tilde{\phi}_k(z_{A_1}, z_{B_1}; \ldots; z_{A_{N/2}},
z_{B_{N/2}})$ this is actually a form well known to physicists. To
see this, imagine a system of $N/2$ bosons where the ``position"
of each boson is specified by two coordinates $(z_1, z_2)$.   The
functions $f_i$ are basis functions for the single ``particle"
positions.  All multiparticle states can be written as symmetrized
(bosonic) linear combinations of the occupancies of these basis
states.


Consider now the case, when we've added $n$ quanta of flux to the
Gaffnian ground state. The highest degree of $f_i()$ is $n$,
and we could choose basis polynomials $f_i(z_1, z_2)$ of the form
$z_1^{n_1} z_2^{n_2}$ with $0 \leq n_1, n_2 \leq n$. The
dimension of this space is  $(n+1)^2$.  However, a different basis
set turns out to be more useful.  Specifically, it is useful to
separate functions $f_i$ that vanish in the limit $z_1 \rightarrow
z_2$, from ones that do not.

We choose a basis for our space of $f_i$ which decomposes into two
disjoint sub-bases: the symmetric $z_1^{n_1} z_2^{n_2} + z_1^{n_2}
z_2^{n_1}$ with $0 \leq n_1 \leq n_2 \leq n$ and the antisymmetric
$z_1^{n_1} z_2^{n_2} - z_1^{n_2} z_2^{n_1}$ with $0 \leq n_1 \leq n_2
\leq n-1$. The dimensions of subspaces spanned by them are
$\frac{1}{2}(n+2)(n+1)$ and $\frac{1}{2}(n+1)n$ respectively.  Clearly
the span of the antisymmetric sub-basis vanishes as $z_1 \rightarrow
z_2$. The quotient of the full space by the span of the antisymmetric
sub-basis is just the span of the symmetric sub-basis ($\mathcal{S}$),
i.e. symmetric polynomials. Of these, polynomials which vanish as $z_1
\rightarrow z_2$ are spanned by $(z_1 - z_2)^2 (z_1^{n_1} z_2^{n_2} +
z_1^{n_2} z_2^{n_1})$, with $0 \leq n_1 \leq n_2 \leq n-2$. The
dimension of this subspace is $\frac{1}{2}n(n-1)$.

The quotient of $\mathcal{S}$ by the subspace of the vanishing
symmetric polynomials has dimension $2n + 1$ and contains
symmetric polynomials in 2 variables that do not vanish in the
limit $z_1 \rightarrow z_2$, we'll call this quotient
$\mathcal{Q}$. However, by considering the Taylor
expansion\cite{ReadRezayiPfaff} of Read-Rezayi pairing form
$\Phi(z_1,z_2)$ given in Eq. \ref{eq:RRpairingphi} we've already
found a set of $2n+1$ linearly independent symmetric polynomials
in $2$ variables, thus we may choose them as the basis of this
quotient space.

Now given a choice of $\tilde \phi_k$ we obtain a zero energy
state of the Hamiltonian.  Further, all possible zero energy
states can be written in this way.   We can now decompose any
$\tilde \phi_k$ into basis polynomials $f_i$ of the above
described form. Let our choice be such that $f_i()$ for $1 \leq i
\leq \frac{F}{2}$ belong to the subspace of polynomials that
vanish as $z_1 \rightarrow z_2$ and $f_i()$ for $\frac{F}{2}+1
\leq i \leq \frac{N}{2}$ belong to the complementary subspace,
i.e. $\mathcal{Q}$.  Then each vanishing $f_i()$ simplifies with
the appropriate factor in the denominator of $\phi_G$ producing a
``broken pair'', and the remaining factors form what we above
called $\Omega()$, whereas the product of non-vanishing $f_i()$
can be reexpressed as a linear combination of Read-Rezayi pairing
forms $\Phi()$. Thus we conclude that the most general zero energy
eigenstate of $H_G$ is of the conjectured form, and therefore we
counted the complete degeneracy of eigenstates for a given value
of additional flux $n$.


\section{An Example of Angular Momentum Addition}
\label{app:explicit}

We would like to determine the full angular momentum spectrum of
the zero energy states of the Hamiltonian $P_3^3$ using Eq.
\ref{eq:decompose}.  Here we will consider the  example of $N=4$
particles and $n=3$ (6 quasiholes). Eq. \ref{eq:decompose} tells
us that we should have a total number of zero energy states given
by the sum of three terms corresponding to $F=0,2,4$.   For $F=4$
we have (6 bosons in 1 orbitals) $\otimes$ (4 fermions in 4
orbitals).  Both 6 bosons in 1 orbital on 4 fermions in 4 orbitals
have $L=0$, so overall this is an $L=0$ state. The $F=2$ case is
more tricky.   Here we have (6 bosons in 2 orbitals) $\otimes$ (2
fermions in 3 orbitals). First we take 6 bosons in 2 orbitals.
When there are two orbitals on a sphere, we clearly have $L=1/2$.
So the two orbitals have $L_z = \pm 1/2$.   There are 7 ways to
fill these orbitals with 6 bosons, which we can write as (6,0);
(5,1), $\ldots$ (0,6). Counting the total $L_z$ of each of these
states, we get $3,2,1,0,-1,-2,-3$ which we recognize as being
$L=3$.  Thus, 6 bosons in 2 orbitals is $L=3$.     Similarly, we
look at 2 fermions in 3 orbitals. The three orbitals must be $L=1$
with $L_z = 1,0,-1$.   We can fill the three orbitals with two
fermions in 3 ways, which have $L_z = 1,0,-1$ so we recognize this
as $L=1$. Now we must add together the angular momentum of (6
bosons in 2 orbitals) $\otimes$ (2 fermions in 3 orbitals).  This
means we need to add $L=3$ with $L=1$.  By the usual angular
momentum addition rules we obtain $L=2,3,4$.     Finally, we turn
to the $F=0$ case.  Here we have (6 bosons in 3 orbitals)
$\otimes$ (0 fermions in 2 orbitals). The 0 fermions in 2 orbitals
clearly has $L=0$.   It is a simple exercise to count up the
possibilities for 6 bosons in 3 orbitals. We discover that this
has $L=0,2,4,6$. Putting together all of the results we find that
the zero energy states of the Hamiltonian $P^3_3$ for $N$
particles with $n=3$ occur at angular momentum $L=0,0,2,2,3,4,4,6$
which agrees with the results of exact diagonalizations.

\vspace*{10pt}

\section{Further Generalized Wavefunctions}

\label{app:further}

Although there may be many possible ways to generalize Gaffnian
wavefunction\cite{endnote2}, the form written in
Eq.~\ref{eq:gaffwf} suggests a generalization from paired to
clustered wavefunctions where instead of dividing the particles
into two groups, we divide the particles into $g$-groups.   Let us
assume the number of particles $N$ in the system is divisible by
$g$ and write $N=g n$. We then write the wavefunction
\begin{widetext}
\begin{equation}
\label{eq:eds} \Psi = \tilde S \left[ \left\{    \prod_{a=1}^{g}
\left[\prod_{(a-1)n < i < j \leq a n} (z_i - z_j)
    \right] \right\} \left\{ \prod_{1 \leq a < b \leq g}
    \left[\prod_{i=1}^{n} \frac{1}{z_{(a-1)n+i} - z_{(b-1)n + i}}
    \right] \right\} \prod_{1\leq i<j \leq N} (z_i - z_j)^{m}
    \right]
\end{equation}
\end{widetext}
with $m \geq 1$ where again $\tilde S$ symmetrizes or
antisymmetrizes for bosons (odd $m$) or fermions (even $m$)
respectively.   Counting powers of $z$ we discover that this
wavefunction occurs at flux \begin{eqnarray}
    N_{\phi}  &=& (N/g -1) - (g-1) + m(N-1) \\
        &=& (1/g+m) N - (g + m)
\end{eqnarray}
corresponding to a filling fraction $\nu=g/(g m + 1)$ which is
just the Jain sequence.  Furthermore the precise value of the flux
(the shift) is also in agreement with the Jain series.  This
construction clearly reproduces the Gaffnian for $g=2$.  For the
bosonic case ($m=1$)  for general $g$ this construction produces a
wavefunction that does not vanish when $g$ particles come to the
same point, but vanishes as $g+1$ powers as the $g+1$'st particle
arrives at that point.  However, for $g > 2$ this trial
wavefunction is not the densest possible wavefunction with this
particular property\cite{Us}.  Nonetheless, we believe that this,
and other related wavefunctions can generally be constructed with
simple projection rules.  For example, for the $g=3, m=1$ case of
Eq. \ref{eq:eds} this wavefunction is the unique densest
wavefunction that does not vanish as 3 particles come together,
that always vanishes as at least 4 powers when 4 particles come
together, and vanishes faster than 4 powers if particles are
brought together in groups of 2 and then two groups of 2 are
brought together.


\begin{references}
\bibitem{endnote2} Another way to write $\Psi$ would be
to use the form of Eq.~2.16 of Ref.~\cite{ReadRezayi} with $k=2$
modifying the function $\chi$ to have the form $\chi(z_1,z_2; z_3,
z_4) = (z_1 - z_3)^2 (z_2 - z_4)^2 (z_1 - z_4) (z_2 - z_3)$. Many
generalizations of this form to clusters rather than pairs are
possible (analogous to appendix \ref{app:further}).
\bibitem{endnote4} Numerically we define the hierarchy 2/5 state
as the ground state of a $V_1$ only Hamiltonian.  For system sizes
for which composite fermion trial wavefunctions can be generated,
the hierarchy and composite fermion wavefunction have near perfect
overlap\cite{Olle}.
\end{references}
\end{document}